\documentclass{article}

\usepackage{arxiv}

\usepackage[utf8]{inputenc} 
\usepackage[T1]{fontenc}    
\usepackage{hyperref}       
\usepackage{url}            
\usepackage{booktabs}       
\usepackage{amsfonts}       
\usepackage{nicefrac}       
\usepackage{microtype}      
\usepackage{lipsum}		
\usepackage{graphicx}
\usepackage{natbib}
\usepackage{doi}

\title{Rapid Antigen Detection Tests Empowered by Cholesteric Spherical Reflectors for Expanded Capabilities}

\author{ \href{https://orcid.org/0000-0000-0000-0000}{\includegraphics[scale=0.06]{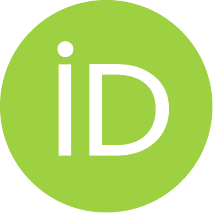}\hspace{1mm}Chih-Chieh Chien}\thanks{These authors contributed equally.}\\
	Department of Physics \& Materials Science\\
	University of Luxembourg\\
	1511 Luxembourg, Luxembourg\\
	\texttt{chihchieh.chien@uni.lu} \\
    \And
    \href{https://orcid.org/0000-0000-0000-0000}{\includegraphics[scale=0.06]{orcid.pdf}\hspace{1mm}Yan-Song Zhang$^*$}\\
	Department of Physics \& Materials Science\\
	University of Luxembourg\\
	1511 Luxembourg, Luxembourg\\
	\texttt{yansong.zhang@uni.lu} \\
    \And
    \href{https://orcid.org/0000-0000-0000-0000}{\includegraphics[scale=0.06]{orcid.pdf}\hspace{1mm}Jean-François Conrotte}\\
	Luxembourg Centre for Systems Biomedicine\\
	University of Luxembourg\\
	4362 Esch-sur-Alzette, Luxembourg\\
	\texttt{jean-francois.conrotte@uni.lu} \\
    \And
    \href{https://orcid.org/0000-0000-0000-0000}{\includegraphics[scale=0.06]{orcid.pdf}\hspace{1mm}Carole L. Linster}\\
	Luxembourg Centre for Systems Biomedicine\\
	University of Luxembourg\\
	4362 Esch-sur-Alzette, Luxembourg\\
	\texttt{carole.linster@uni.lu} \\
    \And
    \href{https://orcid.org/0000-0000-0000-0000}{\includegraphics[scale=0.06]{orcid.pdf}\hspace{1mm}Sivashankar Krishnamoorthy}\\
	Luxembourg Institute of Science \& Technology\\
	L-4362 Esch-sur-Alzette, Luxembourg\\
	\texttt{sivashankar.krishnamoorthy@list.lu} \\
	\And
	\href{https://orcid.org/0000-0000-0000-0000}{\includegraphics[scale=0.06]{orcid.pdf}\hspace{1mm}Jan P.F. Lagerwall} \\
	Department of Physics \& Materials Science\\
	University of Luxembourg\\
	1511 Luxembourg, Luxembourg\\
	\texttt{jan.lagerwall@lcsoftmatter.com} \\
}

\renewcommand{\shorttitle}{\textit{arXiv} Template}

\hypersetup{
pdftitle={Rapid Antigen Detection Tests Empowered by Cholesteric Spherical Reflectors for Expanded Capabilities},
pdfsubject={q-bio.NC, q-bio.QM},
pdfauthor={Chih-Chieh Chien, Yan-Song Zhang, Jean-François Conrotte, Carole L. Linster, Sivashankar Krishnamoorthy, Jan P.F. Lagerwall},
pdfkeywords={Rapid Antigen Test, Cholesteric liquid crystals, Cholesteric Spherical Reflectors},
}

\begin{document}
\maketitle

\begin{abstract}
	While Rapid Antigen Detection Tests (RADTs) have proved their value in point-of-care detection of respiratory diseases, their poor-contrast readout limits their use at early stages of infection where viral concentrations are low, and most RADTs test only for one disease, leaving a sick patient ignorant in case of a negative result. We explore the potential of Cholesteric Spherical Reflectors (CSRs) as enhanced optical labeling probes for RADTs, leveraging their circularly polarized reflections to maximize contrast by background subtraction and thereby lower the analyte concentration required for detection. Moreover, we demonstrate how the ease in tuning the CSR reflection color and the freedom to choose between right- and left-handed circularly polarized reflection allow us to design CSRs to distinguish optically between multiple diseases, eventually enabling RADTs capable of screening for a set of multiple diseases in a single test. We develop a method for functionalizing CSRs with appropriate antibodies and we propose a new type of RADT specifically designed for use with CSRs as detection label, demonstrating that the comparatively large size of CSRs (tens of microns in diameter) does not prevent viral analytes to tether them to an antibody-coated test surface. The strength of the optical CSR signal quantitatively reflects the antigen concentration. 
\end{abstract}


\section{Introduction}
Rapid and accessible detection of respiratory pathogens is important for timely clinical decision-making, infection control, and at-home disease management. Rapid antigen detection tests (RADTs) have become a widely used and valuable tool because they can provide test results of acceptable reliability for many important diseases \citep{Nairz2025} within minutes without requiring specialized equipment. Their analytical performance, however, depends not only on molecular recognition but also on how effectively a target binding event can be converted into an optically detectable signal. Most conventional RADTs rely on colloidal gold, colored latex beads or other chromogenic labeling probes for which detection is based on the appearance and intensity of a visible test line  \citep{Mirica2022}. At low analyte concentrations, the test line can be very faint and hard to distinguish from the background, making the result difficult to assess with certainty. This is further aggravated by the poor optical contrast between the pink or light blue signal color and the white background of standard lateral flow immunoassay RADTs. Optical labeling probes that produce a more distinct and easily recognizable signal, with high contrast of signal against a black background, may therefore provide an attractive strategy for improving readout performance in rapid diagnostic assays.

We argue that such a labeling probe may be provided by Cholesteric Spherical Reflectors (CSRs), beads or shells formed by polymerizing a cholesteric liquid crystal (CLC) precursor \citep{Agha2022a}. CLCs spontaneously develop long-range order with an intrinsic periodic twisted structure that turns CLCs into photonic crystals that selectively reflect circularly polarized light within a narrow and well-defined wavelength range (structural color) \citep{Lagerwall2026, Guo2026,Zhang2022d}. CSRs are made by constraining CLCs to spherical confinement with controlled boundary conditions and then polymerizing them such that the characteristic photonic properties are retained in mechanically stable beads. Their strong wavelength- and polarization-selective reflection gives each CSR a distinct optical signature that can be readily distinguished from the surrounding background \citep{Geng2021}. These properties have enabled CSRs to be used in optical encoding, anti-counterfeiting and many other applications \citep{Avsar2026,Concellon2023,Schwartz2021a,Agha2022a,belmonte2020brush,Kim2023}. Multiple CSRs in a plane develop a highly complex multicolored optical microscopy signature due to photonic cross communication \citep{noh2014tuneable}, so for simplicity we often distinguish different CSRs by their \textit{retroreflection} color \citep{Agha2022a}. This corresponds to the greatest possible reflection wavelength $\lambda_0$ (as measured at the center of the reflection band), occurring for reflection back along the direction of illumination \citep{Geng2021,Agha2022a}.

Beyond these applications, the high optical contrast in detecting individual CSRs thanks to their circularly polarized reflections also enables their use as discrete optical reporters, with each CSR contributing a directly detectable sensing signal. Because we can easily tune the CSR reflection color across the visible spectrum, and we can equally easily choose between right- and left-handed circularly polarized reflection, CSRs enable not only background subtraction but also excellent contrast between different CSR signals. For instance, in a RADT context, a right-handed red-retroreflecting CSR could signify the presence of one antigen, easily distinguishable from a left-handed green-retroreflecting CSR that is used to signal that a different antigen is present. Because of the continuous tunability of retroreflection wavelength and the choice between two orthogonal circular polarizations, CSRs could be used to detect a plethora of different analytes, in principle even in one and the same RADT. This would dramatically expand the capability and versatility of RADTs, which could thus become exceptionally powerful in effective management of disease spreading. However, no attempts have so far been made in functionalizing CSRs with antibodies, as required to use them in RADTs. Here we take on this challenge and explore the potential of CSRs in biosensing by designing, fabricating and assessing a new RADT that uses CSRs as reporters of biological binding events.

\begin{figure}[t!]
	\centering
	\includegraphics[width=\textwidth]{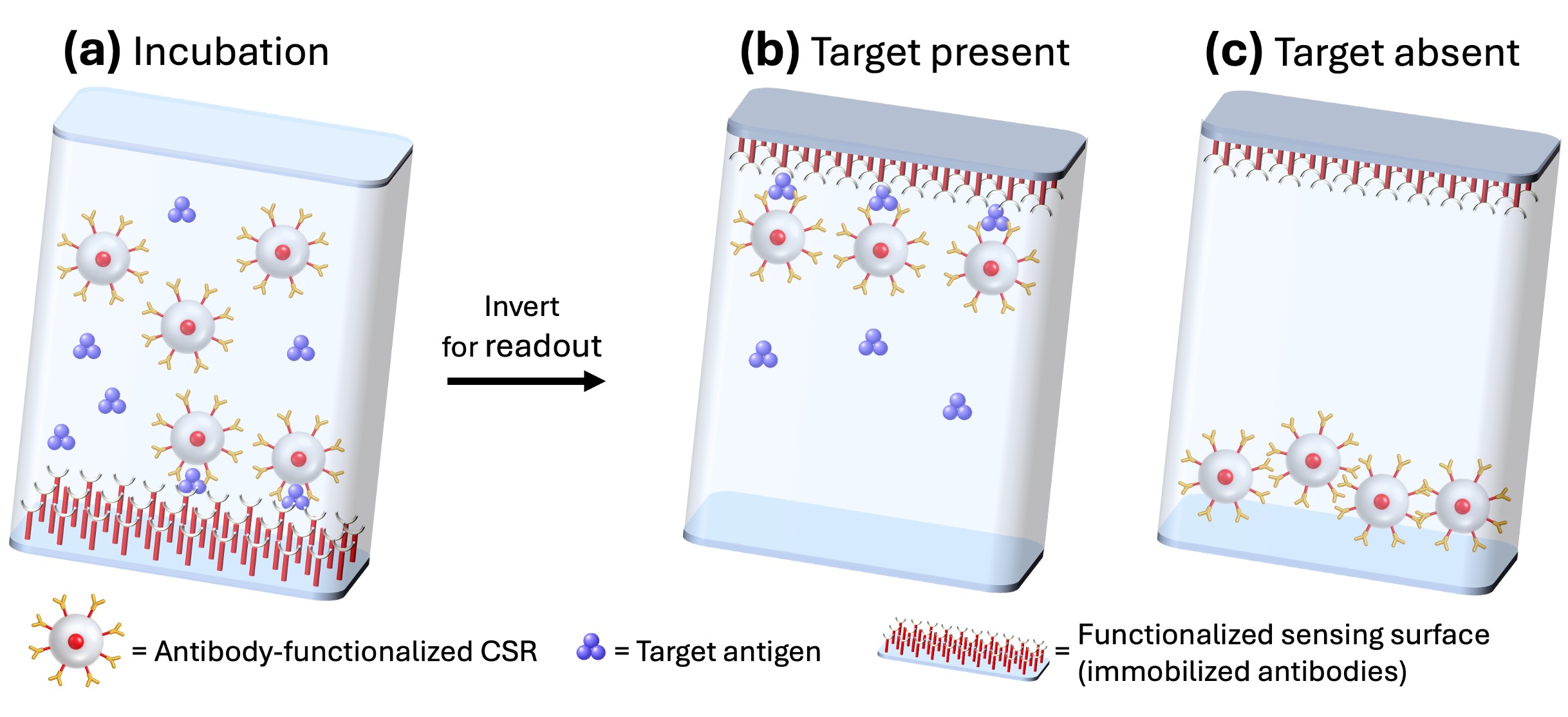}
	\caption{\textbf{Schematic illustration of the gravity-assisted CSR retention assay as a Rapid Antigen Test (not to scale).} (a) During incubation, analyte molecules, when present, bind to recognition elements on the functionalized vial bottom and CSR surfaces, thereby mediating CSR tethering to the vial bottom, functioning as the sensing surface. (b) Following vial inversion for readout, a positive test result is seen from the tethered CSRs at the sensing surface, now facing the user and producing a directly detectable optical signal. (c) In the absence of analyte, CSRs are untethered and therefore move down from the sensing surface under gravity, resulting in little or no CSR signal at the sensing surface constituting the negative test result.}
	\label{fig:fig1}
\end{figure}

The sensing concept is based on converting molecular recognition into an action that physically tethers CSRs to a sensing surface, such that they will deliver a clearly distinguishable optical signal for direct readout, see Fig.\ref{fig:fig1}. Their circular polarization-selective optical response further enables background-suppressed imaging \citep{Geng2021,Agha2022a} and straightforward quantification of the retained CSRs. In this assay, the bottom surface of a transparent vial and the CSR surface are both functionalized with target-binding recognition elements, allowing the analyte to bridge the two surfaces through specific molecular interactions. Because CSRs are slightly denser than the surrounding aqueous medium, inversion of the closed vial provides a simple gravity-assisted separation step: untethered CSRs move away from the sensing surface, whereas analyte-tethered CSRs remain tethered to the surface that is now at the top for visual inspection. The assay response is therefore determined by the CSRs retained at the sensing surface after vial inversion, allowing molecular recognition to be converted into a directly observable optical readout.

To verify this sensing principle in a simple and well-defined recognition system, we first use the high-affinity biotin–NeutrAvidin interaction as a model assay. We then apply the same assay format to SARS-CoV-2 and influenza A (H1N1) analytes, functionalizing the vial bottom and CSR surface with pairs of antibodies directed against different regions of the corresponding viral antigens. The viral antigen experiments confirm antigen-dependent CSR tethering, while concentration-dependent measurements with H1N1 further establish the feasibility of using the number of tethered CSRs as a quantitative optical readout. Together, these results establish CSRs as a powerful alternative to currently dominating latex beads or gold nanoparticles as labeling probes in rapid antigen detection tests, offering numerous unique new opportunities.

\section{Results and Discussion}
\label{sec:R&D}

\begin{figure}[t!]
	\centering
	\includegraphics[width=0.85\textwidth]{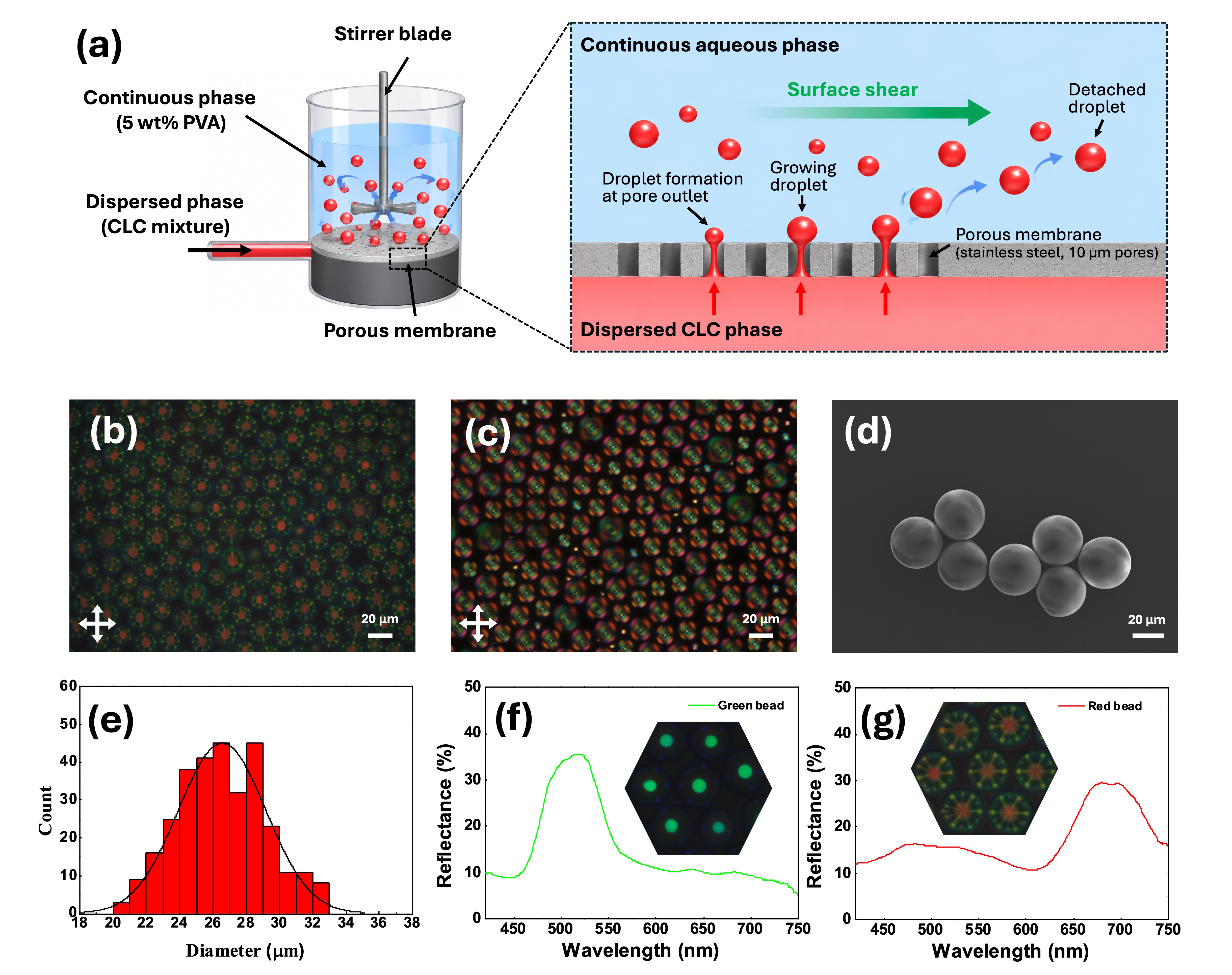}
	\caption{\textbf{Fabrication and optical characterization of solid CSRs.} (a) Schematic representation of the membrane emulsification process used for CSR droplet generation. (b, c) POM images of unpolymerized CSR droplets with red retroreflection observed in reflection mode (b) and transmission mode (c) between crossed polarizers. (d) SEM image of CSR beads after UV photopolymerization. (e) Size distribution of the polymerized CSR beads. (f, g) Center-spot reflection spectra of green-retroreflecting (f) and red-retroreflecting (g) CSRs dispersed in water.}
	\label{fig:fig2}
\end{figure}

\subsection{Preparation and optical characterization of solid CSR beads}
Solid CSR beads with tunable structural color are prepared by membrane emulsification of reactive cholesteric liquid-crystal mixtures at room temperature, as illustrated schematically in Fig.~\ref{fig:fig2}a. The reflected color, quantified by $\lambda_0$, can easily be continuously tuned by adjusting the concentration of the mixture component referred to as chiral dopant; see the Supporting Information, Fig.~\ref{fig:figS1} for all chemical structures and Table~\ref{t1} for quantitative mixture compositions. Following membrane emulsification, the CLC droplets are annealed at room temperature to promote equilibration of the desired cholesteric structure with radial twist axis \citep{Agha2022a}. After about 1~h we confirm a sufficiently well-aligned structure using polarizing optical microscopy (POM), as recognized by the characteristic CSR cross communication pattern in reflection \citep{noh2014tuneable} and a more complex pattern in transmission dominated by optical rotation near the CSR center and birefringence near the perimeter \citep{Geng2016}, see Fig.~\ref{fig:fig2}b–c. The aligned droplets are subsequently photopolymerized to form mechanically stable CSR beads. Scanning electron microscopy (SEM) imaging shows that the polymerized beads retain the spherical morphology with a smooth outer surface (Fig.~\ref{fig:fig2}d).

The CSR beads produced following this procedure have an average diameter of approximately 27~$\mu$m, ranging from 20–33~$\mu$m (Fig.~\ref{fig:fig2}e). This size range provides a practical balance between annealing time of the radial cholesteric configuration and ability to record the optical response of individual CSRs even with relatively simple optical equipment. A small reduction in bead diameter and reflection wavelength is observed after photopolymerization, consistent with polymerization-induced shrinkage. Our CSRs are thus many orders of magnitude larger than the sub-micron particles typically used in commercial RADTs \citep{Mirica2022}; nevertheless, as we will show below, the binding strength of antigen--antibody interactions is strong enough to tether CSRs to our test sensing surface, as required.

The optical properties of the polymerized red- and green-retroreflecting CSRs in water are further examined by reflection POM and optical spectroscopy, see Fig.~\ref{fig:fig2}f–g. Both types of CSRs retain a bright central reflection spot corresponding to their respective $\lambda_0$, confirming that their structural-color and retroreflective properties are preserved after polymerization and transfer to an aqueous environment. These results establish the red- and green-retroreflecting CSR formulations as an optical platform for the subsequent surface-functionalization and sensing studies.

\subsection{Biofunctionalization and validation of recognition-mediated CSR tethering and gravity-assisted differentiation}
\begin{figure}[h!]
	\centering
	\includegraphics[width=0.85\textwidth]{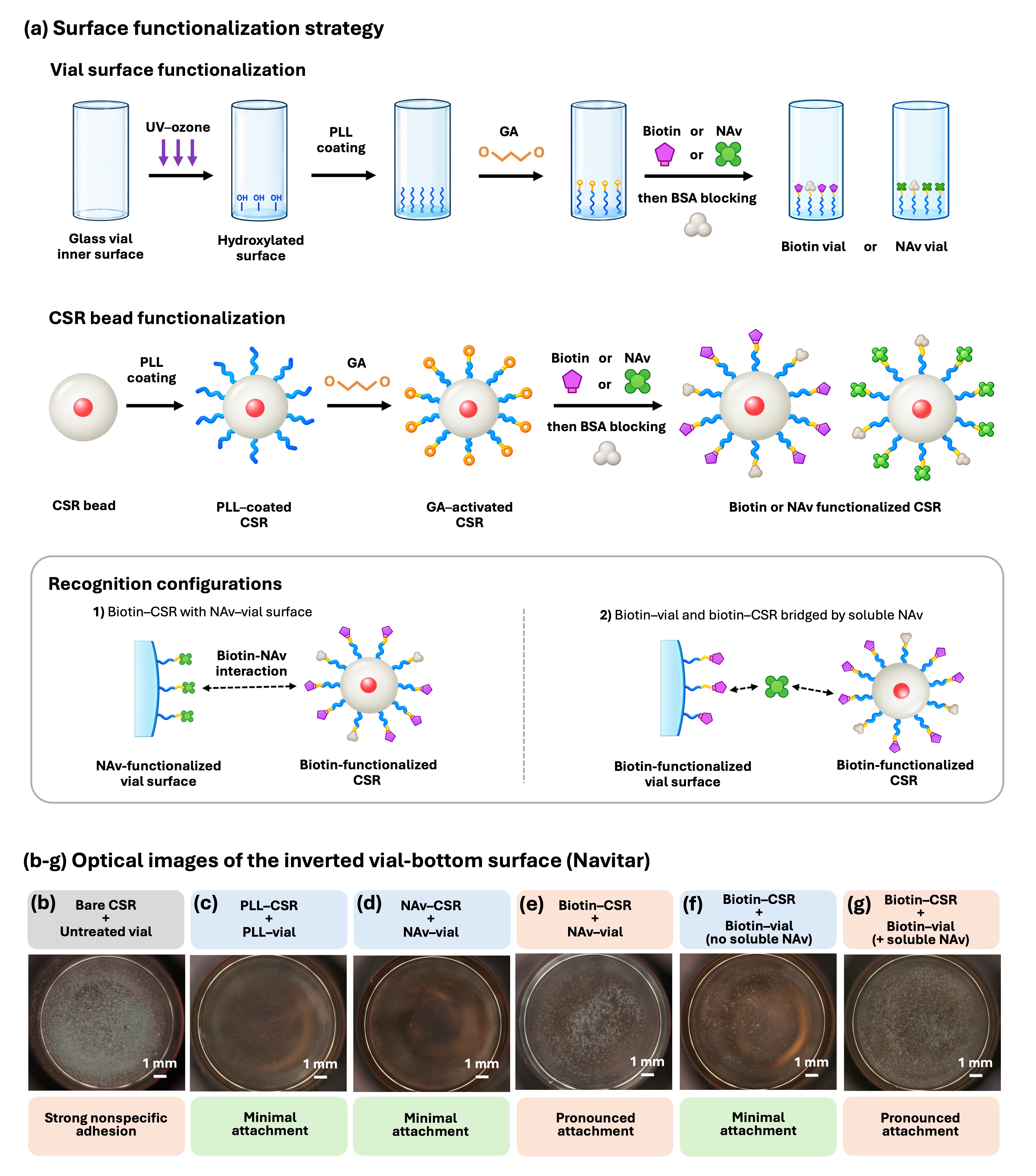}
	\caption{\textbf{Surface biofunctionalization and validation of recognition-mediated CSR tethering.} (a) Schematic illustration of PLL–GA-based functionalization of vial and CSR surfaces with biotin or NAv, followed by BSA blocking. The lower schematics illustrate direct biotin–NAv-mediated attachment and sequential NAv-mediated bridging between biotin-functionalized surfaces. Surface ligand densities are schematic and do not represent quantitative surface coverage. (b–g) Representative optical images of CSR attachment under different surface-functionalization conditions after vial inversion: (b) unmodified CSR and untreated vial; (c) PLL-coated CSR and PLL-coated vial; (d) NAv-functionalized CSR and NAv-functionalized vial; (e) biotin-functionalized CSR and NAv-functionalized vial; (f) biotin-functionalized CSR and biotin-functionalized vial without prior incubation with NAv; and (g) biotin-functionalized CSR and biotin-functionalized vial following incubation of the vial with NAv in solution. Scale bar, 1 mm (applies to b–g).}
	\label{fig:fig3}
\end{figure}

To introduce molecular recognition capability into the CSR-based sensing system, we modify both the CSR beads and the inner surface of the glass vial following a poly-L-lysine (PLL)–glutaraldehyde (GA) functionalization strategy (Fig.~\ref{fig:fig3}a). The glass vial surface is first activated by UV–ozone treatment, followed by sequential modification with PLL and GA, whereas the CSR beads are modified directly using the same PLL–GA chemistry without UV–ozone pretreatment. Biotin or NeutrAvidin (NAv) is subsequently immobilized on the modified surfaces according to the experimental configuration, followed by blocking with Bovine Serum Albumine (BSA) to minimize nonspecific surface interactions. Detailed procedures for vial and CSR surface functionalization are provided in the Methods section.

We evaluate the success of the surface modification procedure using different vials prepared with a variety of bottom surface treatments, incubating them with different sets of CSRs, each with a different surface treatment, suspended in phosphate buffer saline (PBS) as the liquid surrounding medium. After 30 minutes with the CSRs sedimented to the bottom the vial is turned upside down and the fraction of CSRs tethered to the bottom sensing surface (now the top-most surface with respect to gravity) is evaluated optically. We use a low-magnification digital microscope (Navitar) which allows the imaging of the entire test surface. At such low magnification, and with illumination from the side, the red-retroreflecting CSRs appear with a greenish blue color when tethered to the test surface (see Fig.~\ref{fig:fig3}b--g).

As reference we first test for nonspecific CSR adhesion to the test surface. Bare CSRs exhibit substantial adhesion to untreated glass vial surfaces (Fig.~\ref{fig:fig3}b). In contrast, markedly less CSR presence is observed at the sensing surface when both the CSR and vial surfaces are coated with PLL (Fig.~\ref{fig:fig3}c). This reduction in background adhesion provides a more suitable surface condition for subsequent experiments, in which CSR presence at the test surface is intended to arise only by tethering via the defined biomolecular interactions rather than via nonspecific interactions with the vial surface.

As a first confirmation that such biomolecular interactions can indeed keep the large CSRs tethered to the glass vial surface against the action of gravity we explore different configurations of biotin–NAv interactions for mediating the CSR attachment. When both the CSR and vial surfaces are functionalized with NAv and no biotin is present in the system, little attachment is observed, as shown in Fig.~\ref{fig:fig3}d. In contrast, biotin-functionalized CSRs show pronounced attachment to NAv-functionalized vial surfaces (Fig.~\ref{fig:fig3}e), consistent with specific biotin–NAv binding. When both surfaces are functionalized with biotin, little attachment is observed in the absence of NAv (Fig.~\ref{fig:fig3}f). Following incubation of the biotin-functionalized vial with NAv as a free analyte and subsequent addition of biotin-functionalized CSRs, substantial CSR tethering to the vial surface becomes evident (Fig.~\ref{fig:fig3}g). This result supports a sequential mechanism in which surface-bound NAv bridges the two biotinylated surfaces when binding sites remain accessible \citep{Dubacheva2017}. Together, these control experiments confirm recognition-mediated CSR tethering to a properly functionalized glass test surface and provide the basis for the subsequent gravity-assisted retention assay.

\subsection{Establishment of the gravity-assisted quantitative CSR retention assay}
Having established recognition-mediated CSR attachment in the control experiments, we next examine whether NAv-mediated binding can be translated into a concentration-responsive CSR-tethering readout. The biotin–biotin condition without NAv, shown in Fig.~\ref{fig:fig3}f, serves as our background reference for nonspecific CSR retention. Biotin-functionalized vials are first incubated with PBS buffer with increasing concentrations of NAv. After removal of the NAv solution, biotin-functionalized CSRs are introduced and allowed to interact with NAv captured on the vial surface. The vials are then inverted, allowing free CSRs to move away from the sensing surface under gravity while specifically tethered CSRs remain retained at the vial bottom (which is now at the top).

\begin{figure}[t!]
	\centering
	\includegraphics[width=0.85\textwidth]{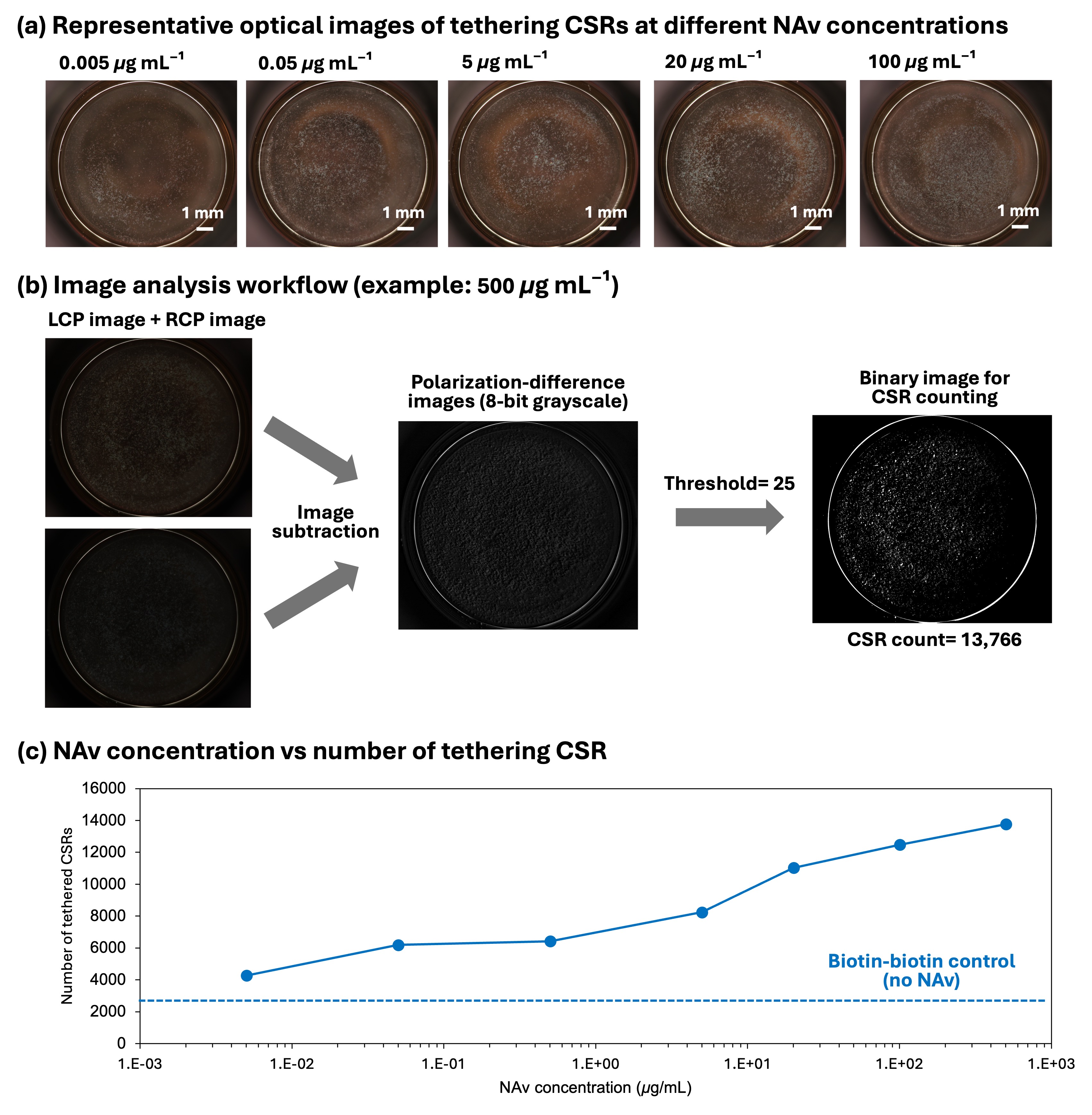}
	\caption{\textbf{Validation of concentration-responsive CSR tethering using NAv as a model analyte.} (a) Representative optical images of CSR tethering to the test surface at different NAv concentrations. (b) ImageJ-based analysis workflow for CSR signal quantification using circular-polarization-difference imaging and threshold-based binary segmentation. (c) Tethered CSR number as a function of NAv concentration (the latter with logarithmic scale). The dashed line represents the biotin–biotin control without NAv.}
	\label{fig:fig4}
\end{figure}

Representative optical images confirm an increasing density of tethered CSRs with increasing NAv concentration, see Fig.~\ref{fig:fig4}a. To quantify this response, we first enhance the contrast by removing the background following the circular-polarization-difference imaging workflow shown in Fig.~\ref{fig:fig4}b and described in the Methods section. Subtraction of images acquired under opposite circular polarizations suppresses common background features, thereby enhancing the circularly polarized CSR signal, enabling threshold-based segmentation and automated CSR counting. The resulting counts are used to construct the concentration-response profile shown in Fig.~\ref{fig:fig4}c.

The number of retained CSRs increases consistently across the tested NAv concentration range, from 4,270 at 0.005~$\mu$g~mL$^{-1}$ to 13,766 at 500~$\mu$g~mL$^{-1}$. The response appears to be nonlinear, with relatively modest changes at the lowest concentrations, followed by an increase that scales roughly linearly with the logarithm of the concentration towards the high-concentration regime. This behavior is consistent with a multivalent retention mechanism rather than a one-to-one relationship between NAv molecules and CSR beads. Because the curvature of the CSRs is negligible at the scale of individual biotin or NAv molecules, each CSR can form multiple biotin–NAv interactions within the bead–vial contact region, hence stable tethering resisting vial inversion is likely to be due to the cumulative tethering strength generated by multiple molecular binding events. At low surface densities of captured NAv, rather few CSRs would then be expected to be sufficiently strongly tethered to the test surface for stable retention upon vial inversion, whereas increasing NAv surface density increases the probability of multivalent tethering. At higher concentrations, the response may become increasingly constrained by the finite number of accessible binding sites, the number of available CSR beads and the vial surface area. These results establish that molecular recognition can be translated into gravity-resisting CSR tethering, also enabling quantitative analysis, providing the mechanistic basis for the subsequent antigen-detection experiments.

\subsection{Viral antigen sensing using antibody-functionalized, dye-doped red and green CSRs}
\begin{figure}[h!]
	\centering
	\includegraphics[width=\textwidth]{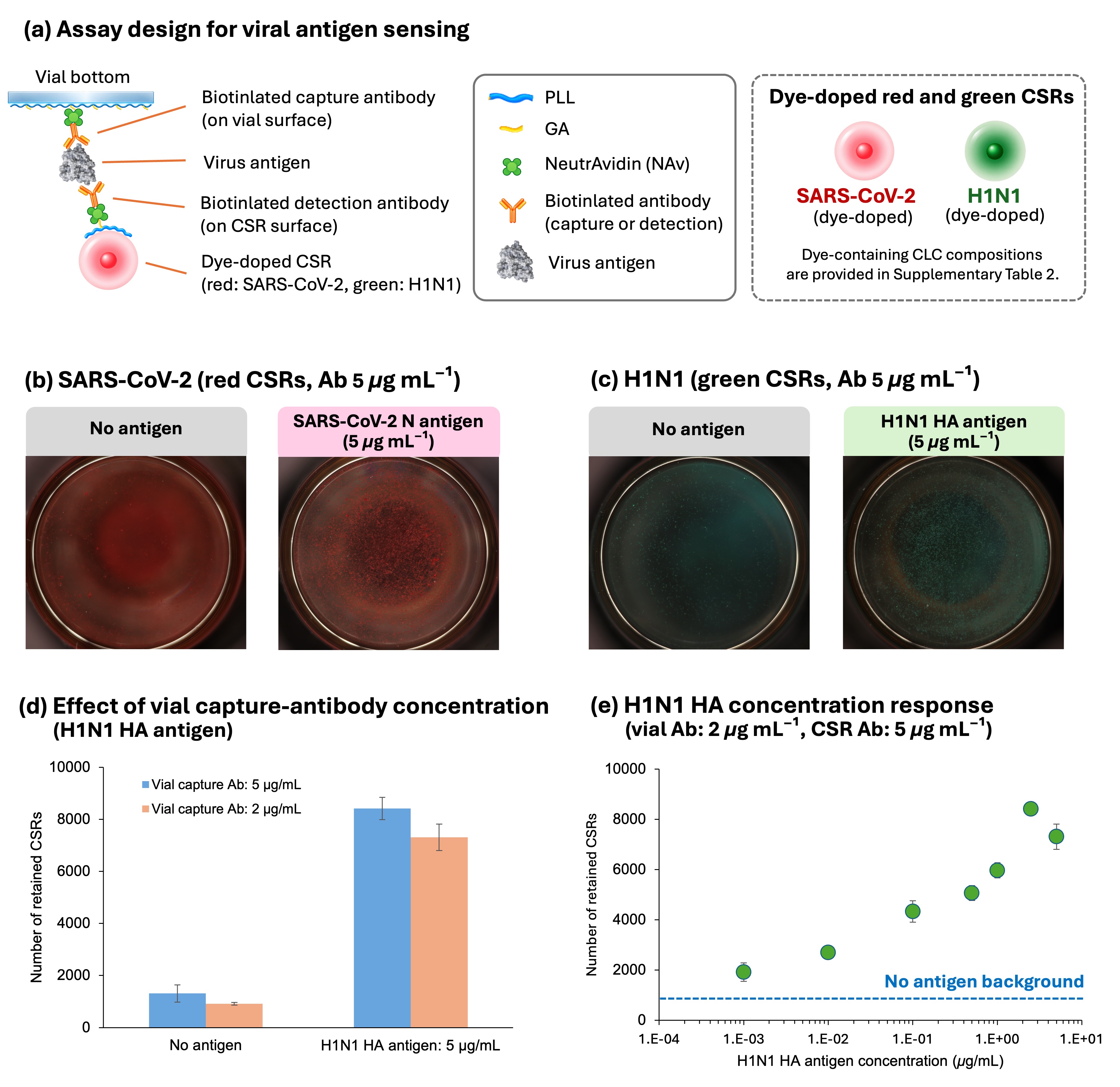}
	\caption{\textbf{Viral antigen sensing using antibody-functionalized d-CSRs.} (a) Schematic illustration of the antibody-functionalized d-CSR retention assay, in which target antigen mediates bridging between capture antibodies on the vial surface and detection antibodies on the CSR surface. (b) Representative optical images showing no red d-CSR tethering in the absence of SARS-CoV-2 nucleocapsid antigen and distinct tethering when the antigen is present at 5~$\mu$g~mL$^{-1}$. Note that the background is always red due to the impact of the dye from the untethered d-CSRs that are not in focus. (c) Representative optical images of the corresponding tests with antibody-functionalized green d-CSRs in the absence and presence of 5~$\mu$g~mL$^{-1}$ H1N1 hemagglutinin antigen, respectively. Again the background color is impacted by the dye in the d-CSRs out of focus. (d) Effect of reducing the vial-surface capture-antibody concentration from 5 to 2~$\mu$g~mL$^{-1}$ on H1N1-mediated CSR retention, while maintaining the CSR detection-antibody concentration at 5~$\mu$g~mL$^{-1}$. Bars represent the mean of two measurements, and error bars indicate the minimum–maximum range ($n = 2$). (e) Number of tethered green CSRs as a function of H1N1 HA antigen concentration using 2~$\mu$g~mL$^{-1}$ capture antibody on the vial surface and 5~$\mu$g~mL$^{-1}$ detection antibody on the CSRs. Points represent the mean of two measurements, and error bars indicate the minimum–maximum range ($n = 2$). The dashed line indicates the no-antigen background.}
	\label{fig:fig5}
\end{figure}

Building on the surface-functionalization strategy and gravity-assisted differentiation mechanism established above, we next configure the CSR platform for viral antigen detection. For each target analyte, the vial and CSR surfaces are functionalized with a pair of biotinylated antibodies directed against different regions of the corresponding viral antigen. The capture antibody is immobilized on the vial surface whereas the detection antibody is immobilized on the CSR surface, in both cases mediated through NeutrAvidin (NAv)–biotin binding. During testing, this allows the target antigen to bridge the two antibody-functionalized surfaces to generate the optical readout signaling a positive test result, see Fig.~\ref{fig:fig5}a. We use red-retroreflecting CSRs for detection of a representative SARS-CoV-2 protein, whereas green-reflecting CSRs are used for detection of a protein specific to influenza A (H1N1), providing optically distinguishable CSR populations for the two target analytes.

Although the red- and green-reflecting CSRs are spectrally distinguishable, strong scattering from the spherical CSRs in water (which is not index-matched to the CSRs) results in relatively pale coloration under direct macroscopic observation. We therefore incorporate absorptive dyes into the corresponding CLC mixtures during CSR preparation to enhance macroscopic color contrast, following a strategy previously established in our laboratory \citep{Avsar2026}. The resulting dye-doped CSRs (d-CSRs) exhibit more distinct red and green coloration while preserving the polarization-selective optical response used for subsequent background subtraction and image analysis. The compositions of the d-CSR precursor mixtures are provided in Supplementary Table~\ref{t2}.

Antibody immobilization is achieved by extending the PLL–GA functionalization scheme established above. Following PLL and GA modification, NAv is immobilized on both the vial and CSR surfaces, followed by BSA blocking. Biotinylated capture and detection antibodies are then attached to the vial and CSR surfaces, respectively, through the high-affinity biotin–NAv interaction. A second BSA blocking step is subsequently performed to further reduce nonspecific adhesion. For the initial proof-of-concept experiments, both the vial and CSR surfaces are functionalized using 5~$\mu$g~mL$^{-1}$ of the corresponding biotinylated antibodies. The use of Biotin tethered to the surface using a long and flexible PEG tether is expected to enhance efficiency of capture of the NAv, and potentially, the biotinylated antibodies to the NAv \citep{Beggiato2023}. Detailed functionalization procedures are described in the Methods section.

The antigen-responsive behavior of the antibody-functionalized platform is first evaluated by comparing conditions with and without target antigen. For the SARS-CoV-2 system, only limited tethering of red CSRs is observed in the absence of antigen, whereas pronounced CSR tethering is observed in the presence of 5~$\mu$g~mL$^{-1}$ SARS-CoV-2 nucleocapsid (N) antigen (Fig.~\ref{fig:fig5}b). A similar antigen-dependent response is observed for influenza A H1N1 using green CSRs, with greater CSR tethering in the presence of 5~$\mu$g~mL$^{-1}$ recombinant H1N1 hemagglutinin (HA) antigen than in the corresponding no-antigen control (Fig.~\ref{fig:fig5}c). These results demonstrate that antigen-mediated recognition between the antibody-functionalized vial and CSR surfaces can be translated into gravity-assisted and CSR-mediated positive/negative differentiation for two different viral targets.

For quantitative comparison of tethered CSR numbers, the circular-polarization-difference imaging workflow established above is applied throughout. Corresponding images acquired under opposite circular polarizations are subtracted to suppress common background (which is here dominated by the dye in untethered d-CSRs which are out of focus) and scattering features, thereby enhancing CSR contrast and facilitating threshold-based segmentation and counting. This background suppression is particularly useful for resolving sparse d-CSRs under weak-tethering conditions.

We next examine the effect of reducing the capture-antibody concentration on the vial-bottom surface while maintaining the CSR functionalization conditions unchanged. In the initial experiments, both surfaces are functionalized using 5~$\mu$g~mL$^{-1}$ of the corresponding antibodies. For subsequent experiments, the detection-antibody concentration used for CSR functionalization is maintained at 5~$\mu$g~mL$^{-1}$, whereas the capture-antibody concentration on the vial surface is reduced from 5 to 2~$\mu$g~mL$^{-1}$. This reduction decreases the absolute number of tethered CSRs in both the absence and presence of antigen. However, the primary objective is not to maximize the total number of tethered CSRs but to maintain clear discrimination between antigen-positive samples and the no-antigen background. Under the reduced vial-surface antibody condition, background retention is lower while antigen-dependent retention remains readily distinguishable from the corresponding control (Fig.~\ref{fig:fig5}d). We therefore select the 2~$\mu$g~mL$^{-1}$ capture-antibody concentration as a practical compromise between antibody consumption, background retention, and preservation of antigen-dependent signal discrimination, while the CSR detection-antibody concentration is maintained at 5~$\mu$g~mL$^{-1}$.

Using this configuration, the number of tethered CSRs shows an overall increase with increasing H1N1 HA antigen concentration over the tested range of 0.0001–5~$\mu$g~mL$^{-1}$ (Fig.~\ref{fig:fig5}e). At the lowest antigen concentrations, CSR tethering remains relatively close to the no-antigen background, whereas stronger tethering is generally observed with increasing antigen concentration. Although some variability is observed at the higher concentrations, the overall response shows a clear concentration-dependent trend. These results demonstrate that antibody-mediated CSR tethering can provide a concentration-responsive optical readout for viral antigen sensing.

\section{Conclusions and Outlook}
We have developed a new type of Rapid Antigen Detection Test (RADT) that uses Cholesteric Spherical Reflectors (CSRs) functionalized with disease-relevant antibodies for detecting and quantifying the presence of viral analytes such as proteins from SARS-CoV-2 and H1N1 Influenza. The concept is simple, based on an antibody-coated bottom surface of a glass vial in which the sample fluid is injected, followed by a small volume of the suspension with antibody-coated CSRs. After an incubation period during which the CSRs are tethered by the antigen to the vial bottom in case of a positive sample the vial is turned upside down. In case of a negative sample the CSRs sediment down to the vial top (now at the bottom) due to the CSR density being higher than that of water. In contrast, the tethering of the CSRs to the vial bottom (now at the top) in case of a positive sample allows a simple high-contrast detection of the CSR signal in the color range and with the circular polarization selected for detection of the disease in question. A quantitative relation between target analyte concentration and the number of tethered CSRs is achieved but the minimum analyte concentration for ensuring a clear signal is still higher than desired. Further optimization of the surface biofunctionalization as well as the details of the RADT operation procedure should allow significant improvement in this respect.  

We have also tested the combination of the red SARS-CoV-2 d-CSRs and the green H1N1 d-CSRs in a single vial coated with the antibody of either viral protein to explore the possibility of combining both tests in a single RADT. At the stage of writing, this combination gives rise to false positives, hence further optimization is needed to ensure specificity, possibly involving a more strategic selection of antibody--antigen combination for each target virus. The results presented above show, however, that CSRs with distinctly different optical properties can be prepared to detect a specific viral antigen, hence we are confident that the false positives are not due to our RADT concept or the use of CSRs as labels, but rather due to our rudimentary surface functionalization process and our limited options in choice of antibodies. With strategic optimization we believe CSR-based RADTs can be designed and fabricated that can successfully screen for multiple diseases during a single test, to the great benefit of patients wishing to understand which disease, out of multiple possible candidates, is affecting them. Such high-information-content RADTs would also be of great value from a societal perspective, not least given the current wave of vaccine skepticism and the risk of new pandemics that may soon need to be managed.

\section{Methods}
\subsection{Fabrication of CLC droplets by membrane emulsification}
CLC droplets were produced at room temperature using a Micropore Technologies Emulsifier LDC-1 membrane emulsification system. The CLC mixture (see the supporting information for composition) was loaded into a syringe and delivered at a flow rate of 0.1~mL min$^{-1}$ through a stainless-steel membrane with 10~$\mu$m pores in a continuous aqueous phase of 50~mg~mL$^{-1}$ poly(vinyl alcohol) (PVA, $M_w = 13$--$23$~kg~mol$^{-1}$,
87--89\% hydrolyzed, Sigma-Aldrich). A controlled shear stress of 18.7~Pa was applied at the membrane surface to promote droplet detachment. Under these conditions, relatively uniform CLC droplets were generated.

\subsection{Annealing and photopolymerization of CSR beads}
Following emulsification, approximately 100 mL of the CLC droplet suspension in aqueous PVA was transferred into a Petri dish with a diameter of 15 cm. The droplets were then incubated at room temperature for 1~h to allow development of the radial twist axis cholesteric configuration. The degree of alignment was evaluated by reflection polarized optical microscopy. After formation of the characteristic radial texture, the droplets were photopolymerized in a UV-curing chamber (Opsytec Dr. Gröbel, BSL-01), with the wavelengths 330-450 nm and light intensity of 200~mW~cm$^{-2}$ for 5~min. The Petri dish was then removed from the UV-curing chamber and gently shaken by hand to redistribute the droplets. The suspension was subsequently exposed to UV light for an additional 5~min to improve irradiation of the densely packed droplets. The resulting solid CSR beads were washed several times with deionized water to remove residual PVA before further use.

\subsection{Optical microscopy and spectral characterization}
The optical textures of the CLC droplets and polymerized CSR beads were examined by reflection and transmission polarizing optical microscopy using an Olympus BX51 microscope. For measurements in aqueous conditions, the solid CSRs were dispersed in water and introduced into flat glass capillaries. Reflection spectra were recorded for green-, and red-reflecting CSR beads, and the central reflection wavelength, $\lambda_0$, was determined from the corresponding spectra.

\subsection{Scanning electron microscopy}
For high-resolution imaging of the surface morphology of dried polymerized CSR beads we used a JEOL JSM-6010LA scanning electron microscopy (SEM). Prior to imaging the CSRs were coated with a thin layer of gold to avoid charge build-up.

\subsection{Surface functionalization of glass vials}
Commercial glass vials (2~mL, VWR, catalog nr. NSCAB7999-1) were used as the sensing substrates. The vials were placed upright in a UV–ozone chamber (Jelight 42-220) and treated for 1~h to activate the glass surface. Immediately after UV–ozone treatment, 150~$\mu$L of 1.0~mg~mL$^{-1}$ poly-L-lysine (PLL, Sigma Aldrich) solution in water was added to each vial, which was closed and then incubated overnight at room temperature. The PLL solution was then removed and the vials were rinsed three times with deionized water.

The PLL-modified surfaces were subsequently treated with 150~$\mu$L of 8~vol.\% glutaraldehyde (GA, diluted from Sigma Aldrich Grade I, 70\% in H$_2$O, purified for use as an electron microscopy fixative) for 1~h at room temperature. The GA solution was removed and the vials were rinsed three times with 0.1× phosphate-buffered saline (PBS, Sigma-Aldrich, 10× liquid concentrate, molecular biology grade), prepared by first diluting the 10× PBS stock solution to 1x, followed by a further tenfold dilution to 0.1×.

For NeutrAvidin functionalization, the GA-activated vials were incubated with 150~$\mu$L of 0.1~mg~mL$^{-1}$ NeutrAvidin (NAv, Fisher Scientific PI31000, purchased in powder form) for 1~h at room temperature to allow GA-mediated immobilization of the protein. For biotin functionalization, the GA-activated vials were incubated with 150~$\mu$L of 2~mM biotin–PEG–amine (Biopharma PEG, purity $\geq$95\%, $M_w \approx 1$~kDa) under the same conditions. Following functionalization, the solutions were removed, and the surfaces were blocked with 150~$\mu$L of 1.0~mg~mL$^{-1}$ bovine serum albumin (BSA) for 30~min at room temperature to minimize nonspecific adsorption. The vials were finally rinsed three times with 0.1× PBS before use.

\subsection{Surface functionalization of CSR beads}
CSR beads were functionalized using the same PLL–GA surface chemistry as that used for the glass vials, except that no UV–ozone treatment was applied. For each functionalization, 10–15~mg of CSR beads were transferred into a 1.5~mL polypropylene microcentrifuge tube. All CSR functionalization steps (described below) were performed under continuous vortex-assisted agitation at room temperature to maintain bead dispersion and ensure uniform contact with the functionalization solutions.

The CSR beads were incubated with 500~$\mu$L of 1.0~mg~mL$^{-1}$ aqueous PLL solution overnight. The beads were then collected by centrifugation in an Eppendorf 5810R centrifuge at 4000~rpm for 90~s, and the supernatant was removed. The beads were washed three times with deionized water; for each wash, the bead suspension was manually shaken for 30~s, followed by centrifugation at 4000~rpm for 90~s and removal of the supernatant.

The PLL-modified beads were subsequently treated with 500~$\mu$L of 8~vol.\% GA solution for 1~h. Following incubation, the beads were collected by centrifugation and washed three times with 0.1× PBS using the same washing procedure.

For NAv functionalization, the GA-activated CSR beads were incubated with 500~$\mu$L of 0.5~mg~mL$^{-1}$ NAv for 1~h. For biotin functionalization, the beads were incubated with 500~$\mu$L of 2~mM biotin–PEG–amine (average $M_w \approx 1$~kDa) for 1~h. After functionalization, the beads were collected by centrifugation and blocked with 500~$\mu$L of 1.0~mg~mL$^{-1}$ BSA solution for 30~min. The functionalized beads were finally washed three times with 0.1× PBS using the same washing procedure before use.

\subsection{Gravity-assisted CSR retention assay}
For the NeutraVidin model assay, biotin-functionalized glass vials were first incubated with 150~$\mu$L NAv solutions at the desired concentrations for 30~min at room temperature. The NAv solution was then removed, and each vial was filled with 2~mL of 1x PBS. A 50~$\mu$L aliquot of the biotin-functionalized CSR suspension was subsequently added to the vial and incubated for a further 30~min at room temperature to allow binding of the CSR biotin to NAv captured on the vial surface. Following incubation, the vial was inverted, allowing untethered CSR beads to move away from the sensing surface under gravity while tethered beads remained at the vial bottom (now at the top). The vial-bottom surface was then imaged through the inverted vial using an optical imaging system (Navitar) equipped with a camera (Pixelink).

\subsection{Antibody functionalization of CSR beads and glass vials}
Following PLL–GA modification as described above, the CSR beads and glass vials were further functionalized with target-specific biotinylated antibodies through NAv–biotin binding. For CSR functionalization, GA-activated beads were incubated with 500~$\mu$L of 0.5~mg~mL$^{-1}$ NAv for 1~h at room temperature under continuous vortex-assisted agitation. The beads were subsequently collected by centrifugation and washed three times with 0.1× PBS using the washing procedure described above. The NAv-modified beads were then blocked with 500~$\mu$L of 1.0~mg~mL$^{-1}$ BSA for 30~min at room temperature. Following washing, the beads were incubated with 500~$\mu$L of 5~$\mu$g~mL$^{-1}$ biotinylated anti-SARS-CoV-2 nucleocapsid protein antibody (Abcam, rabbit monoclonal antibody) or biotinylated anti-influenza A hemagglutinin antibody (antibodies-online, rabbit polyclonal antibody) for 1.5~h at room temperature under gentle continuous rotation at 20 rpm using a Loopster digital tube rotator (IKA). After antibody immobilization, the beads were washed and subjected to a second blocking step with 500~$\mu$L of 1.0~mg~mL$^{-1}$ BSA for 30~min at room temperature under the same gentle rotational mixing conditions. The antibody-functionalized CSR beads were finally washed three times with 0.1× PBS and resuspended in PBS before use.

For vial functionalization, GA-activated vial surfaces were incubated with 150~$\mu$L of 0.1~mg~mL$^{-1}$ NAv for 1~h at room temperature. After removal of the NAv solution, the vials were rinsed three times with 0.1× PBS and blocked with 150~$\mu$L of 1.0~mg~mL$^{-1}$ BSA for 30~min. The NAv-functionalized vial surfaces were subsequently incubated with 150~$\mu$L of 2~$\mu$g~mL$^{-1}$ biotinylated capture antibody against the corresponding SARS-CoV-2 nucleocapsid protein or influenza A H1N1 hemagglutinin protein for 1.5~h at room temperature, followed by three rinses with 0.1× PBS. Surface Plasmon Resonance (SPR) measurements indicated a SARS-CoV-2 antibody surface loading of approximately 6–8~ng~cm$^{-2}$ on the NAv-modified surface, substantially lower than the theoretical monolayer coverage estimated from antibody dimensions. Based on this experimentally observed surface loading, a biotinylated antibody concentration of 2~$\mu$g~mL$^{.1}$ was used for functionalization. A second blocking step was then performed using 150~$\mu$L of 1.0~mg~mL$^{-1}$ BSA for 30~min. The vials were finally rinsed three times with 0.1× PBS before use.

For each target, the capture and detection antibodies were selected as a matched pair recognizing distinct epitopes of the corresponding antigen.

\subsection{SARS-CoV-2 and Influenza A antigen assay}
For SARS-CoV-2 and influenza A antigen detection, 150~$\mu$L of the corresponding antigen solution, either recombinant SARS-CoV-2 nucleocapsid protein (Abcam, catalog no. ab289630, purity $>95\%$) or recombinant influenza A H1N1 hemagglutinin protein (Abcam, catalog no. ab307788, purity $>95\%$), at the desired concentration was added to the antibody-functionalized vial and incubated for 30~min at room temperature. The antigen solution was subsequently removed, and the vial was filled with 2~mL of 1x PBS. A 50~$\mu$L aliquot of CSR beads functionalized with the corresponding detection antibody was then introduced and incubated for 30 min at room temperature. After incubation, the vial was inverted for gravity-assisted separation of untethered CSRs, and the vial-bottom surface was imaged through the inverted vial using the same optical setup.

\subsection{Image analysis and quantification of tethered CSR beads}
The number of CSR beads tethered to the vial bottom was quantified from circularly polarized optical images using ImageJ. For each sample, images were acquired sequentially using left- and right-circular polarizers under identical imaging conditions. To enhance CSR contrast and suppress common background features, a polarization-difference image was generated by subtracting the image acquired under the opposite circular polarization from that acquired under the circular polarization matching the handedness of the corresponding CSR population. The resulting difference image was converted to 8-bit grayscale. For the NAv concentration-dependent retention experiments shown in Fig.~\ref{fig:fig4}, a fixed threshold range of 25–255 was applied, whereas a threshold range of 10–255 was used for the SARS-CoV-2 and H1N1 antigen-sensing experiments shown in Fig.~\ref{fig:fig5}. CSR beads were subsequently identified and counted using the Analyze Particles function in ImageJ. Within each experimental series, identical image-processing parameters were applied to all samples.

\section{Acknowledgments}
We gratefully acknowledge financial support from the Luxembourg National Research Fund (FNR),
grant reference C21/MS/16325006 BIOFLICS (for the purpose of open access, the authors have applied a Creative Commons Attribution 4.0 International (CCBY 4.0) license to any Author Accepted Manuscript version arising from this submission).

\bibliographystyle{unsrtnat}
\bibliography{references}  

\newpage

\textbf{\Large{Supporting Information}}

\renewcommand{\thefigure}{S\arabic{figure}}
\renewcommand{\thetable}{S\arabic{table}}

\renewcommand{\thesection}{S\arabic{section}}
\renewcommand{\refname}{Supplementary References}

\setcounter{figure}{0}
\setcounter{table}{0}
\setcounter{section}{0}
\setcounter{page}{1}

\section{Chemical structures of materials used}\label{chemicals}
The CLC mixtures used to create CSRs of different colors are shown in Table~\ref{t1}. They all contain the bi-functional polymerizable mesogen RM257, the monofunctional polymerizable monomers A6OCB (ST3474), RM23 (ST2670), the nonmesogenic additive 1,6-hexanediol diacrylate (HDDA), the chiral dopant R5011 inducing a right-handed twist in the non-chiral host mixture establisehd by the other components, and the photoinitiator IRG651. The HDDA is added at a concentration of 5.0~wt\% to facilitate room-temperature emulsification. It reduces the room-temperature viscosity from 1085~mPa~s for the corresponding mixture without HDDA to 550~mPa~s, and it lowers the clearing temperature from approximately 72.4 to 45.2$^\circ$C (Fig.~\ref{fig:figS2}). This enables stable processing under ambient conditions.

\begin{figure}[h!]
	\centering
	\includegraphics[width=\textwidth]{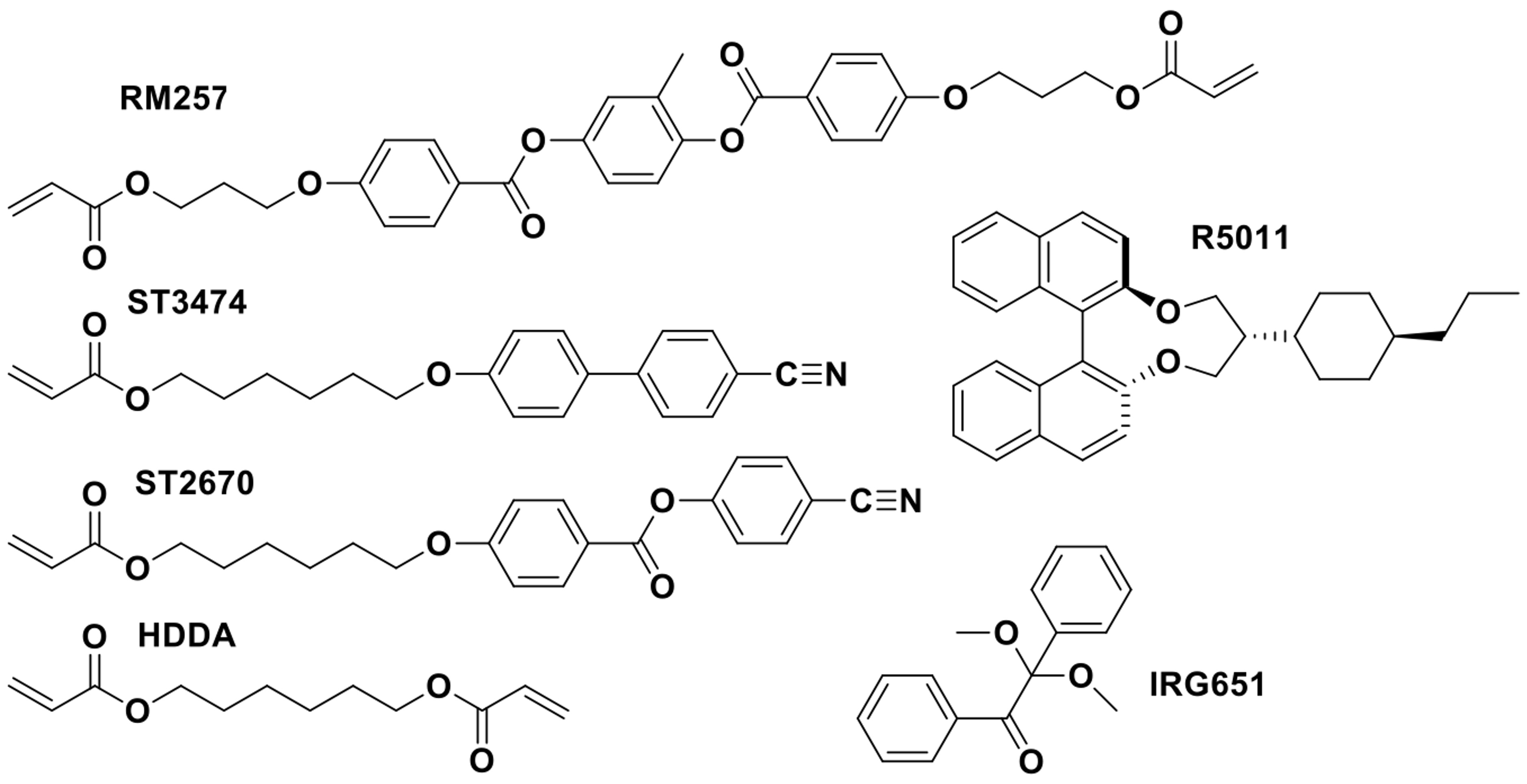}
	\caption{Chemical structures of the CLC components.}
	\label{fig:figS1}
\end{figure}

The exact compositions for red and green retroreflection color of CSRs are given in Table~\ref{t1}. The main difference is the amount of chiral dopant R5011 that is used. By increasing its concentration the retroreflection color shifts to shorter wavelenght, in this case from from red to green.

\begin{table}[h!]
    \caption{Composition of the red- and green-reflecting CLC mixtures used for CSR preparation.}\label{t1}
    \centering
    \begin{tabular}{ccccccc}
     \textbf{CLC formulation} & \textbf{RM257 (wt\%)} & \textbf{A6OCB (wt\%)} & \textbf{RM23 (wt\%)} & \textbf{HDDA (wt\%)} & \textbf{R5011 (wt\%)} & \textbf{Irg651 (wt\%)} \\
    Red CLC & 43.3 & 28.3 & 19.6 & 5.0 & 1.8 & 2.0 \\
    Green CLC & 43.1 & 28.1 & 19.5 & 5.0 & 2.3 & 2.0 \\

    \end{tabular}

    \label{t2}
\end{table}

\begin{figure}[h!]
	\centering
	\includegraphics[width=\textwidth]{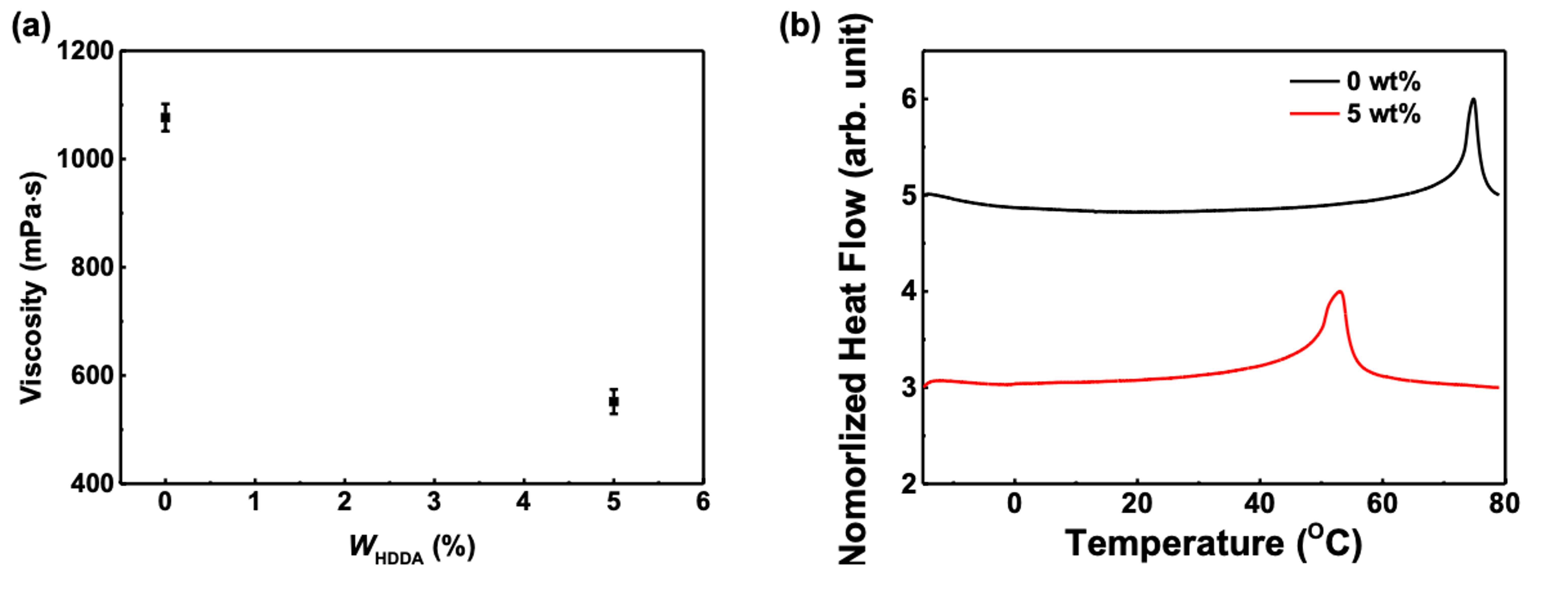}
	\caption{(a) The effective shear viscosity (see text for comments on interpretation) of CLC mixtures as a function of $\omega_{HDDA}$, i.e., the mass percentage of HDDA in the CLC mixture, at room temperature. (b) Differential scanning calorimetry (DSC) thermograms (first heating run, with a rate of 5$^\circ$C/min.) of the crystallized CLC base mixture, without HDDA. The indicated temperatures are onset melting and clearing temperatures. (b) DSC thermograms (second heating run, with a rate of 10$^\circ$C/min.) of CLC mixtures with different HDDA concentrations. The indicated temperatures are the onset temperatures for clearing in each mixture.}
	\label{fig:figS2}
\end{figure}

\begin{table}[h!]
    \caption{Composition of dye-doped red and green CLC mixtures used for CSR preparation}
    \centering
    \begin{tabular}{ccccc}
     \textbf{Formulation} & \textbf{Base CLC} & \textbf{Base CLC content (wt\%)} & \textbf{Dye} & \textbf{Dye content (wt\%)} \\
    Dye-doped red CLC & Red CLC & 99.0 & Oil red dye & 1.0 \\
    Dye-doped green CLC & Green CLC & 99.7 & Dipon transparent black dye & 0.3 \\
    \end{tabular}
    \label{t2}
\end{table}

\end{document}